# EFFECT OF DISSOLVED OXYGEN CONTENT ON PHOTOCATALYTIC PERFORMANCE OF GRAPHENE OXIDE

M. Bakhtiar Azim[a,c*], Farzana Nargis[b], Sajib Aninda Dhar[d], Md. Rakibul Qadir[d], Md. Abdul Gafur[d], A.S.W Kurny[a], Fahmida Gulshan[a]

[a]Department of Materials and Metallurgical Engineering, Faculty of Engineering, Bangladesh University of Engineering and Technology, Dhaka-1000, Bangladesh

[a]Professor [retired], Department of Materials and Metallurgical Engineering, Faculty of Engineering, Bangladesh University of Engineering and Technology, Dhaka-1000, Bangladesh.

[a]Professor, Department of Materials and Metallurgical Engineering, Faculty of Engineering, Bangladesh University of Engineering and Technology, Dhaka-1000, Bangladesh.

[b]Department of Chemical Engineering, Faculty of Engineering, Bangladesh University of Engineering and Technology, Dhaka-1000, Bangladesh

[c]School of Engineering Science, Faculty of Applied Science, Simon Fraser University, Burnaby, British Columbia, Canada.

[d]Pilot Plant and Process Development Center, Bangladesh Council of Scientific and Industrial Research (PP and PDC, BCSIR), Dhaka-1205, Bangladesh.

Corresponding Author: M. Bakhtiar Azim, E-mail: bakhtiar31.mme@gmail.com; mbazim@sfu.ca

*Abstract—* Graphene, a two-dimensional (2D) promising emergent photocatalyst consisting of earth-abundant elements. This study evaluated the potential of graphene oxide (GO) towards photocatalytic degradation of a novel organic dye, Methylene Blue (MB). In this work, photocatalytic activity of graphene oxide (GO), graphene oxide (GO) along with hydrogen peroxide ($H_2O_2$) were tested by photodegrading Methylene Blue (MB) in aqueous solution. The resulted GO nanoparticles were characterized by X-ray powder diffraction (XRD), Scanning Electron Microscopy (SEM) and Energy Dispersive Spectroscopy (EDX) and Fourier Transform Infrared Ray Spectroscopy (FTIR). The XRD data confirms the sharp peak centered at $2\theta \approx 10.44°$ corresponding to (002) reflection of GO. Based on our results, it was found that the resulted GO nanoparticles along with $H_2O_2$ achieved ~92% photodecolorization of MB compared to ~63% for $H_2O_2$ under natural sunlight irradiation at pH~7 in 60 min. The influences of oxygen and hydrogen peroxide ($H_2O_2$) on the degradation of MB during sunlight/GO process were investigated. Experimental results indicated that oxygen was a determining parameter for promoting the photocatalytic degradation. The rate constant of degradation ($k_1$) increased from 0.019 to 0.042 $min^{-1}$ for dissolved oxygen content (DOC) 3.5 $mgL^{-1}$ when direct photocatalysis (MB/GO) and $H_2O_2$-assisted photocatalysis (MB/$H_2O_2$/GO) were used. Owing to the fact that $H_2O_2$ acted as an electron and hydroxyl radicals (•OH) scavenger, the addition of $H_2O_2$ should in a proper dosage to enhance the degradation of MB. Moreover, as the initial concentration of dissolved oxygen (DO) was increased from 2.8 to 3.9 $mgL^{-1}$, the rate constant of degradation ($k_1$) increased from 0.035 to 0.062 $min^{-1}$. The mechanism of photodegradation and kinetics were also studied for both direct photocatalysis and $H_2O_2$-assisted photocatalysis.

## INTRODUCTION

The discharge of azo dyes, which are stable and carcinogenic, into water bodies are harmful to human health, and cause such illness as cholera, diarrhea, hypertension, precordial pain, dizziness, fever, nausea, vomiting, abdominal pain, bladder irritation, staining of skin [1]. Dyes also affect aquatic life by hindering the photosynthesis process of aquatic plants, eutrophication, and perturbation [2,3]. Therefore, numerous techniques have been applied to treat textile wastewater, such as activated carbon adsorption (physical method), chlorination (chemical method), and aerobic biodegradation (biochemical method) [4]. However, further treatments are needed, which create secondary pollution in the environment, such as the breakdown of parent cationic dyes to Benzene, $NO_2$, $CO_2$, and $SO_2$ [5]. Advanced oxidation processes (AOPs) are widely applied to mineralize dyes into $CO_2$ and $H_2O$ [6,7]. AOPs include ozonation, photolysis, and photocatalysis with the aid of oxidants, light, and semiconductors. Photocatalytic degradation was initiated when the photocatalysts absorb photons (UV) to generate electron-hole pairs on the catalyst surface. The positive hole in the valence band ($h_{VB}^+$) will react with water to form hydroxyl radical (•OH), followed by the oxidization of pollutants to $CO_2$ and $H_2O$ [8].

Methylene Blue (MB), also known as Swiss Blue, is an azo dye (Table 1). MB is widely used in textile industries for dye processing, and upto 50% of the dyes consumed in textile industries are azo dyes [8-10]. In the past few years, several catalysts have been used to degrade MB, such as $BiFeO_3$ [4], $TiO_2$ [5], ZnO [8], and ZnS [12], and the results were summarized in (Table 2).

Table 1. Properties of Methylene Blue (MB).

| Properties | Cationic Azo Dye |
|---|---|
| Synonym name | Swiss Blue |
| Molecular formula | $C_{16}H_{18}ClN_3S$ |
| Molecular weight | 319.851 g/mol |
| Absorbance wavelength($\lambda_{max}$) | 664 nm |
| Molecular structure | 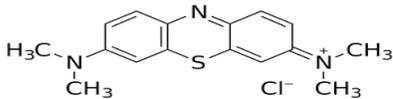 |

Table 2. The photocatalytic degradation of MB using several catalysts.

| Authors/ Year | Catalysts | Degradation efficiency (%) | Conditions | References |
|---|---|---|---|---|
| Soltani et al. 2014 | $BiFeO_3$ | 100% MB | Time: 80 min; Catalyst loading: 0.5 g/L$^{-1}$; Irradiation: Natural Sunlight; pH 2.5 | [4] |
| Dariani et al. 2016 | $TiO_2$ | 100% MB | Time: 2 hr; Catalyst loading: 0.5 g/L$^{-1}$; Irradiation: UV light; pH 2.5 | [5] |

Photocatalysis enabled by graphene-family nanomaterials has received considerable attention in recent years [13,14]. A common strategy to design such photocatalysts is to combine some graphene-family materials with semiconducting materials, such as TiO2, to form nanocomposite photocatalysts. It is believed that this approach promotes the flow of electrons from semiconducting photocatalysts to graphene-related materials upon photoexcitation, thereby inhibiting the electron−hole pair recombination and increasing the photocatalysis efficiency [13,14]. However, in such a system, the hydroxyl radical (•OH) produced during photocatalysis may react with certain graphene materials to result in rapid decomposition of the latter [15-17].

GO, structurally analogous to graphene, possesses an apparent bandgap because of its association with a range of oxygen-containing groups that concomitantly enhances its dispersion into water [18−20]. While GO possesses some interesting properties similar to semiconducting materials [18,21,22] its sole use in photocatalysis has not been well delineated. Recent studies by Yeh et al. demonstrated that GO can photocatalyze the splitting of water to generate a considerable amount of $H_2$ [23−25]. Hsu et al. reported a possible conversion of $CO_2$ to methanol using GO as the photocatalyst [26]. These studies suggest that GO alone may act as a potential photocatalyst. GO as a carbonaceous, metal-free nanomaterial is also attractive in photocatalysis, as it does not involve expensive noble metals frequently used in photocatalytic systems [27−29].

Graphene oxide (GO) has more oxygen functional groups than reduced graphene oxide (rGO) and a surface area of 736.6 $m^2/g$ [31] compared to 400 $m^2/g$ [32] for graphite. Numerous methods have been used in the synthesis of GO, such as chemical, thermal, microwave, and microbial/bacterial [33]. Chemical exfoliation is preferable due to its large-scale production and low cost. Chemical exfoliation involves three steps, oxidation of graphite powder, dispersion of graphite oxide (GTO) to graphene oxide (GO) and GTO exfoliation by ultrasonication to produce graphene oxide (GO) [34]. GO, with its unique electronic properties, large surface area and high transparency, contributes to facile charge separation and absorptivity in its structure. As a potential photocatalytic material, GO has been used in the decolorization of Methylene Blue [38] and Rhodamine B [38].

$H_2O_2$ is a clean oxidant as well as a fuel that generates $O_2$ and $H_2O$ upon decomposition [39,40]. It finds wide applications including fuel cells, organic synthesis, bleaching agents, and advanced oxidation processes (AOPs) such as Fenton reaction ($Fe^{2+}/H_2O_2$), and UV 254 nm/$H_2O_2$ for generating •OH for pollution removal and disinfection [39,40].

Advanced oxidation processes (AOPs) have attracted wide interests in waste water treatment since 1990s. It is widely applied to mineralize dyes into $CO_2$ and $H_2O$ by generating •OH due to their high oxidation potential. AOPs include ozonation, photolysis, and photocatalysis with the aid of oxidants, light (sunlight specially UV light) and semiconductors. Photocatalytic degradation is initiated when the photocatalysts absorb photons (hv) to generate electron-hole pairs on the catalyst surface. The positive hole in the valence band ($h_{VB}{}^+$) will react with water to form hydroxyl radical (•OH), followed by the oxidization of pollutants to $CO_2$ and $H_2O$.

Moreover, it has been reported that GO can efficiently photocatalyze the generation of $H_2O_2$ to millimolar levels under simulated sunlight in a few hours. The concentration of $H_2O_2$ produced is among the greatest values reported in current photocatalytic systems without organic electron donors. Hou et al. showed that dissolved $O_2$ played a pivotal role in the photoproduction of $H_2O_2$ by GO and that superoxide ($O_2•^−$) was not involved and the results indicate that GO is a promising, metal-free photocatalyst to generate H2O2 in an environmentally sustainable manner [41].

In this investigation, a facile method to prepare GO nanoparticles has been reported which were synthesized via chemical oxidation. The photocatalytic performances of the prepared GO and GO with $H_2O_2$ were evaluated in the degradation of a model organic dye, methylene blue (MB) in aqueous solution under sunlight. To best of our knowledge, detailed investigations on catalyst loading, initial dye concentration, and initial solution pH are still lacking. This study aims to determine the optimum experimental conditions for the best photodecolorization performance.

## EXPERIMENTAL SECTION

*Chemicals and Materials*

Graphite powder and Sodium Nitrate were purchased from Sigma Aldrich (Steinheim, Germany). Sulfuric acid (98%) was obtained from Merck (Darmstadt, Germany). Potassium permanganate, Hydrochloric acid (37%) and Hydrogen Peroxide (30%) were also purchased from Sigma Aldrich (Steinheim, Germany). The chemicals were used without further purifications. Methylene Blue (MB) powder from Sigma-Aldrich (Steinheim, Germany) was used as the model compound in this study. Deionized water was used throughout the experiments.

*Synthesis of Graphene Oxide (GO)*

Graphene oxide was produced by the modified Hummers' method by oxidizing the graphite powder [21]. In a typical synthesis, 3g of graphite powder and 1.5g $NaNO_3$ were mixed with 69 ml $H_2SO_4$ (conc. 98%) in a beaker. Then, 9g of $KMnO_4$ was slowly added and stirred in an ice-bath for 1 h below 20°C. Then the mixture was heated to 35°C and kept stirring for 2 hrs. Then, an oil bath was maintained at a temperature of 95°C~98°C . After that the beaker was placed in the oil bath for 15 minutes and 150 ml Deionized water was added slowly while stirring. After cooling the mixture to room temperature, again an oil bath was set at a temperature of 60°C and maintained and the beaker was kept in the oil bath for additional 60 minutes at a constant temperature of 60°C. Then 150 ml Deionized water was slowly added in the beaker while stirring. Finally, dropwise addition of 30 ml (30%) $H_2O_2$ was made and stirred for 2 hrs. Then washing, filtration and centrifugation were performed until removal of $Cl^-$ ions by using Deionized water. Finally, the resulting precipitate was dried at 70°C for 24 hrs in an oven giving thin sheets which was Graphite Oxide (GTO). Graphite Oxide was made into a fine powder form by grinding in a crucible and then GTO powder was finely dispersed in Deionized Water. Then, ultrasonication was carried out for the complete exfoliation of GTO to GO.

*Characterization & Analytical Method*

The X-ray diffraction pattern of GO was recorded by a Bruker, D8 Advance diffractometer (Germany). The sample was scanned from 5° to 80° using Cu K$\alpha$ radiation source ($\lambda$ = 1.5406 A°) at 40 kV and 30 mA with a scanning speed of $0.01°s^-1$. The surface morphology of GO was observed by FESEM-JEOL (FEG-XL 30S) Field Emission Scanning Electron microscope (FESEM). FTIR spectra of GO was recorded by a Agilent Cary 670 FTIT spectrometer. The photodegradation percentage of MB was determined by using an ultraviolet-visible spectrophotometer (Shimadzu-UV-1601) at $\lambda_{max}$ = 664 nm and wavelength region between 400 and 800 nm. DI water was used as a reference material. The DO concentration was quantified by an oxygen membrane electrode (Oxi 320, WTW).

*Photocatalytic Reaction*

Photocatalytic experiments were carried out by photodegrading MB using UV-Vis spectroscopy. The solution of MB (pH~7) without GO was left in a dark place for 60 min. Then, the dye solution was exposed to sunlight irradiation and there was no decrease in the concentration of dye. In a typical experiment, 7.5 mg of GO was added into a 100 mL 0.05 mM MB solution. Before illumination, the suspensions were continuously stirred at dark place for 60 min to reach an adsorption-desorption equilibrium between the photocatalyst and MB. Then, the suspensions were exposed to sunlight irradiation for another 60 mints and samples were taken at regular time intervals (0 min, 10 min, 15 min, 30 min, 45 min, and 60 min) and filtered to remove the GO. Where required, the initial pH of solution (pH~7) was adjusted by small amount of 0.1 M NaOH and 0.1 M HCl. Photodegradation was also observed for 0.05 mM MB solution using only $H_2O_2$ and GO along with $H_2O_2$. The decolorization efficiency of MB was determined by using the equation shown below:

**Photodegradation efficiency (%) = $[(C_0 - C_t) / C_0] \times 100\%$**  (1)
$$= [(A_0 - A_t) / A_0] \times 100\%$$
**[According to 'Beer-Lambert Law']**

where, $C_0$ is the initial concentration of MB, $C_t$ is the concentration of MB at time, t and $A_0$ is the initial absorbance of MB, $A_t$ is the absorbance of at time, t.

### III. RESULTS AND DISCUSSIONS

*Characterization of GO*

The powder X-ray diffraction pattern of GO shows a broadened diffraction peak (Fig 1(a)) at around $2\theta \approx 10.44°$, which corresponds to the (002) reflection of stacked GO sheets.

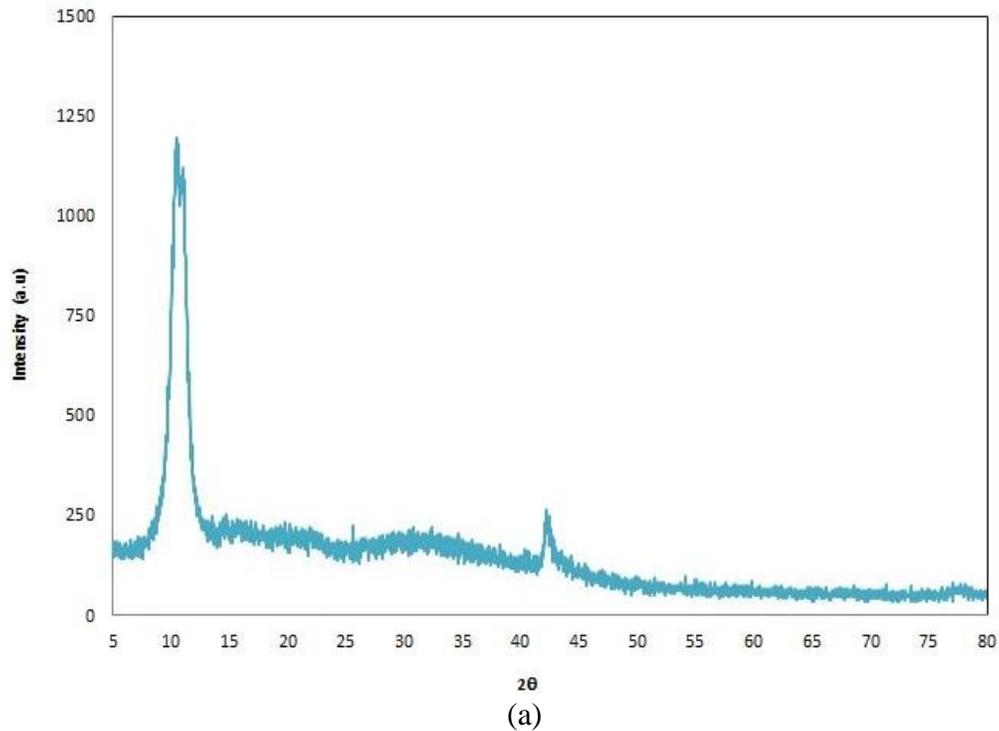

(a)

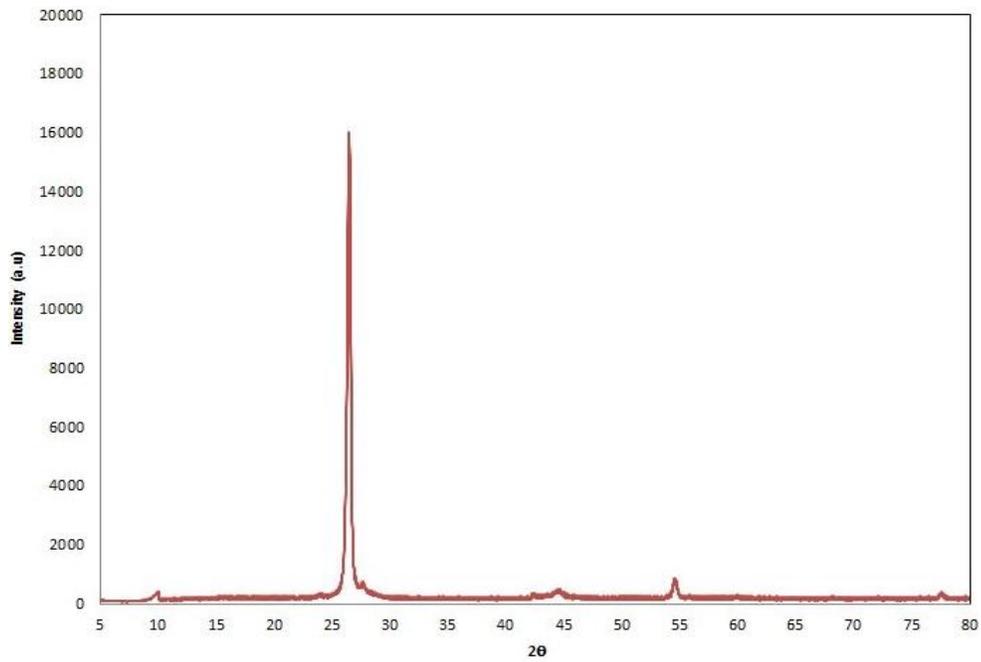

(b)

Fig 1: XRD pattern of (a) Graphene Oxide (GO), and (b) Graphite powder.

SEM images of GO structure with different magnifications are shown in (Fig 2). SEM images of GO shows the crumbled sheet of GO layers.

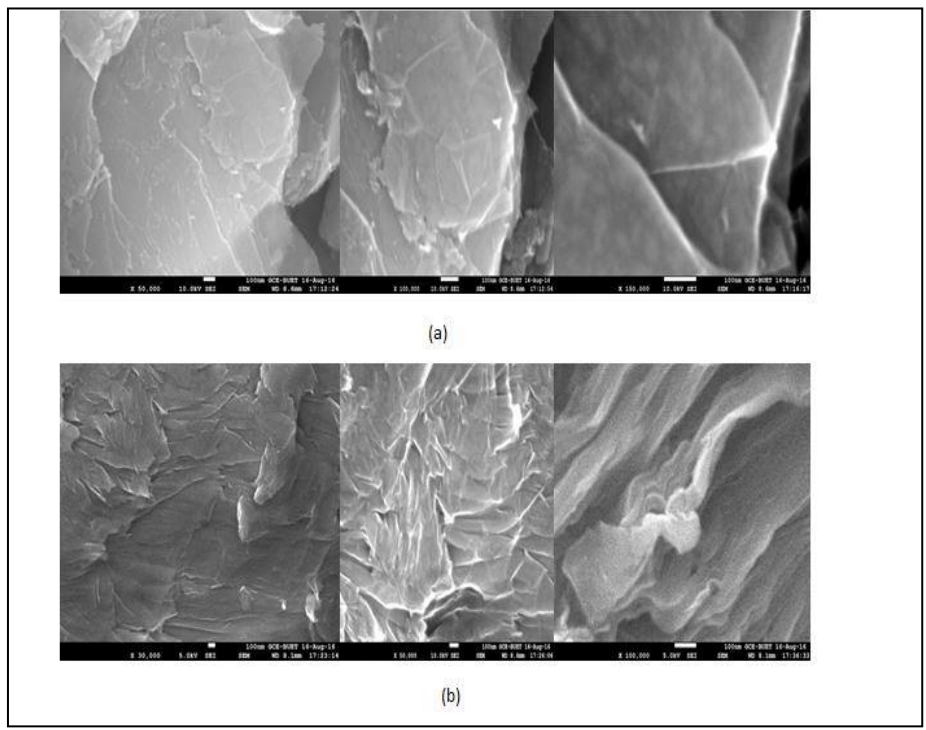

Fig 2: SEM Image of (a) Graphite powder and (b) Graphene Oxide (GO).

EDX studies are generally carried out to test the elemental composition and purity of the sample by giving us the details of all the elements present in the given sample. The EDX spectra and elemental composition of GO is shown in (Fig 3).

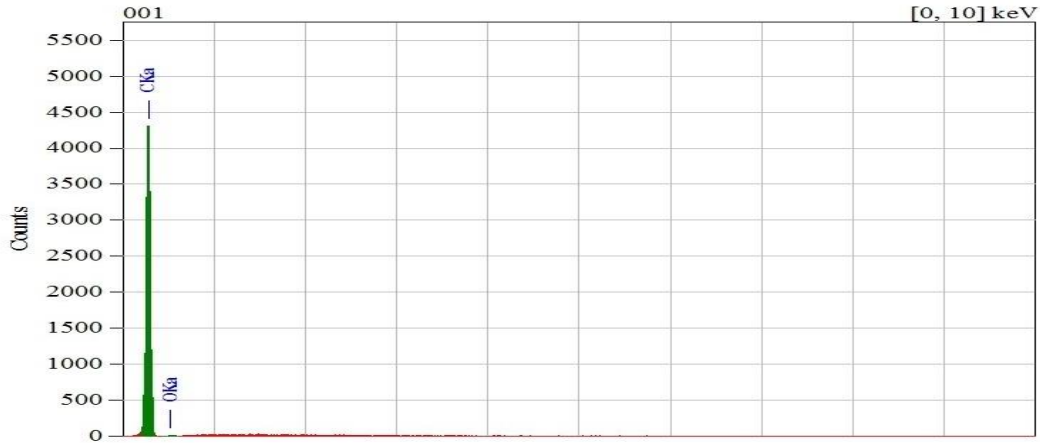

(a)

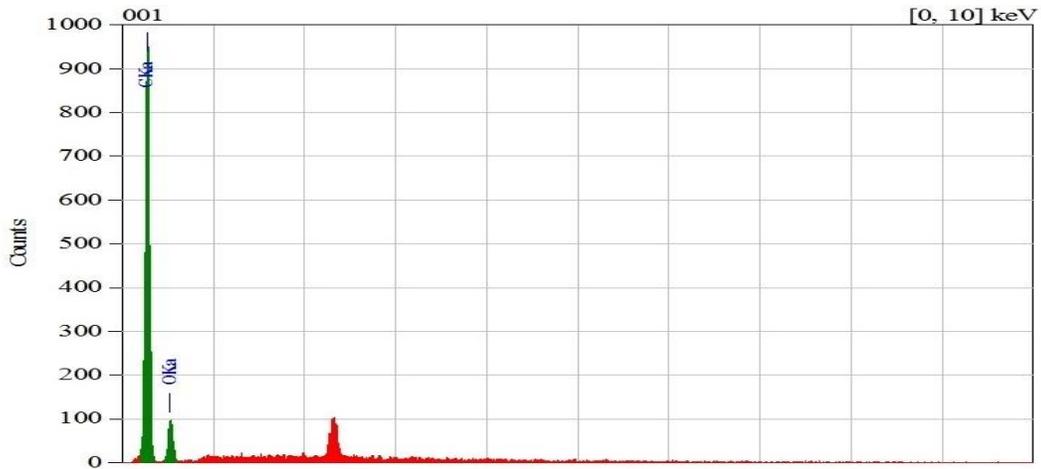

(b)

Fig 3: EDX spectra of (a) Graphite powder, and (b) Graphene Oxide (GO).

FTIR analysis of GO shows broad absorption spectrum observed at ~3420 cm$^{-1}$ corresponding O-H stretching vibration indicating existence of absorbed water molecules and structural O-H groups in GO. The broad peak appeared in GO spectrum depicted the presence of O-H & C-H stretching. Besides, a band at 1747 cm$^{-1}$ might be related to not only the C=O stretching motion of COOH groups situated at the edges and defects of GO lamellae but also that of ketone or quinone groups. The peak near 1700-1550 cm$^{-1}$ widens and moves to 1565 cm$^{-1}$ that reflects the presence of un-oxidized aromatic regions (Fig 4).

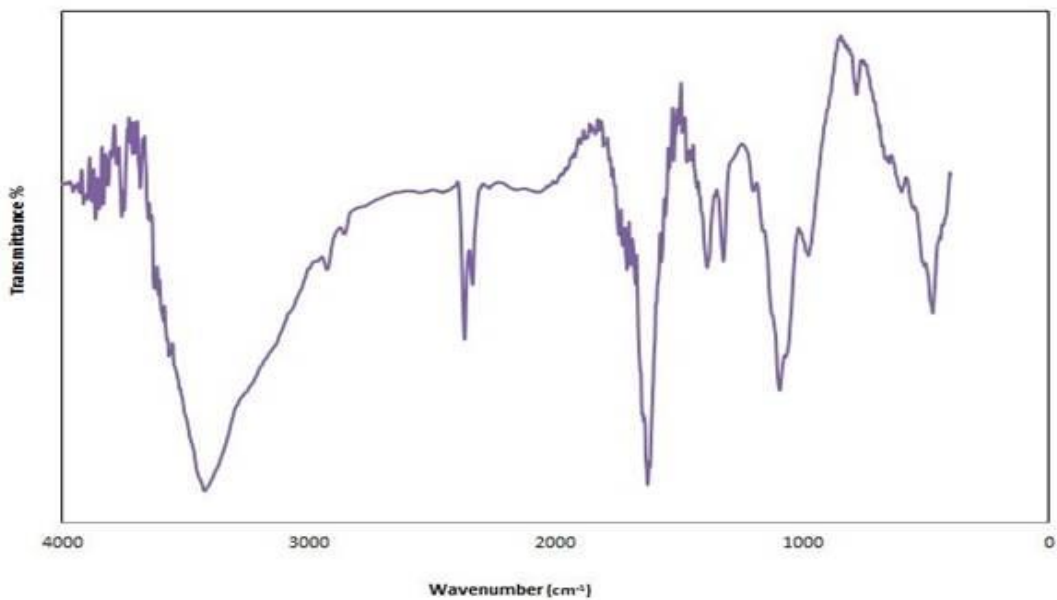

Fig 4: FTIR Spectra of Graphene Oxide (GO).

*Photocatalytic Activity of GO*

The photocatalytic activity of GO was evaluated by measuring the photodegradation of MB as a function of irradiation time under natural sunlight. MB, having intense absorption at 664 nm. The solution was stirred well and allowed to natural sunlight irradiation at regular intervals and the corresponding absorption spectra were measured. MB dye (0.05 mM) was diluted in 100 ml DI water. The photo catalytic degradation of MB was studied after addition of 7.5 mg of GO in 6 ml $H_2O_2$ to the 100 ml dye solution using sonication. Irradiation was carried out in volumetric flask under the sunlight. UV-Vis was used to measure absorbance of the dye solution at regular time intervals. Controlled experiments were also carried out to confirm the degradation of MB by UV-Vis. Experiments were repeated for only $H_2O_2$ and for only GO. Under natural sunlight irradiation GO along with $H_2O_2$ showed 92.23% photodegradation efficiency after 60 min whereas only GO and only $H_2O_2$ showed 68.68% and 62.81% respectively (Fig 5).

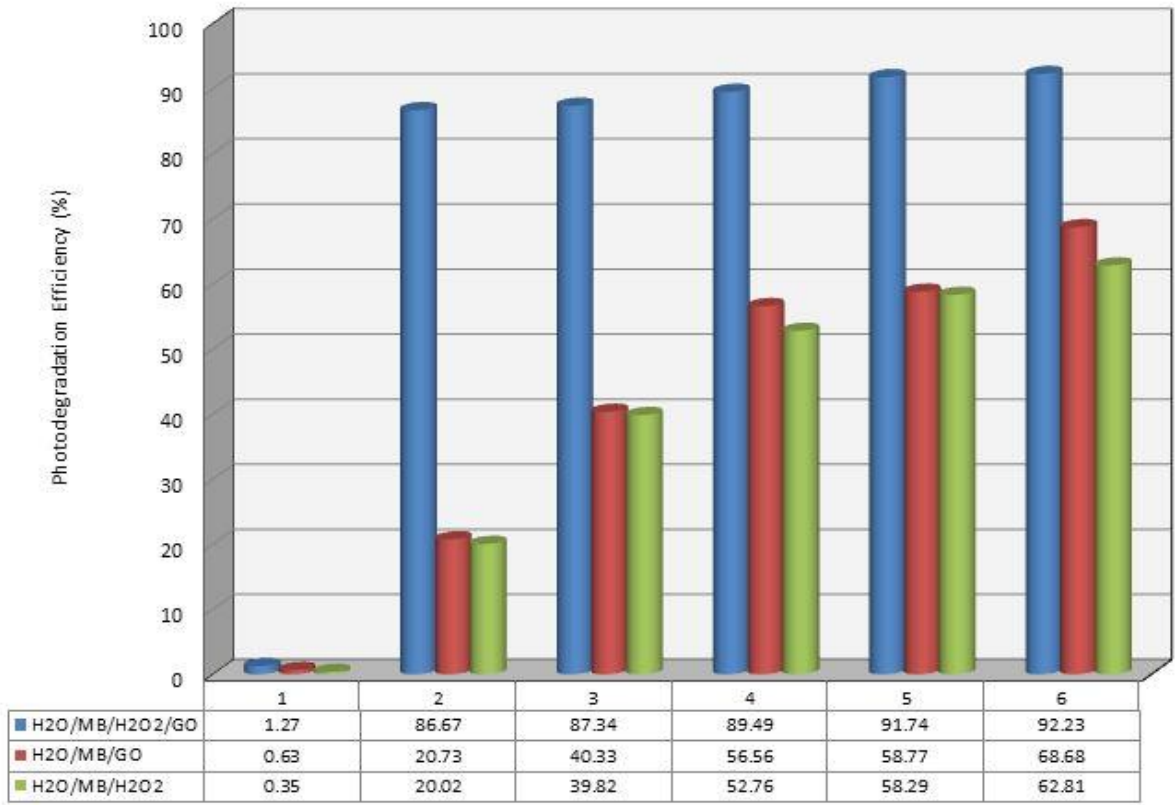

Fig 5: Photodegradation efficiency of the $H_2O/MB/GO$, $H_2O/MB/H_2O_2$, and $H_2O/MB/H_2O_2/GO$.

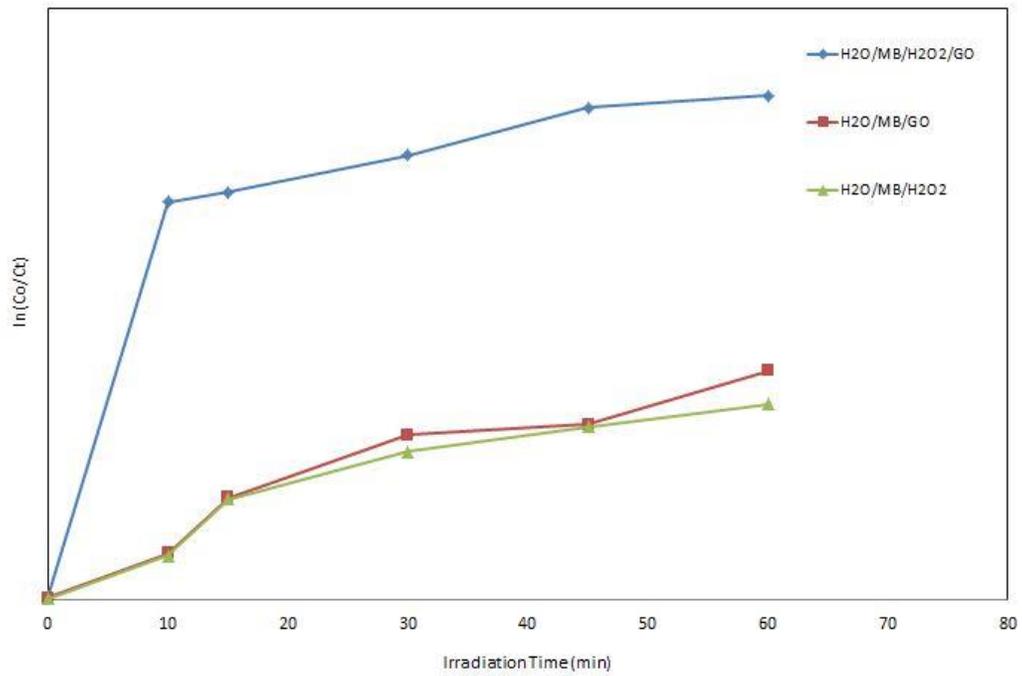

(a)

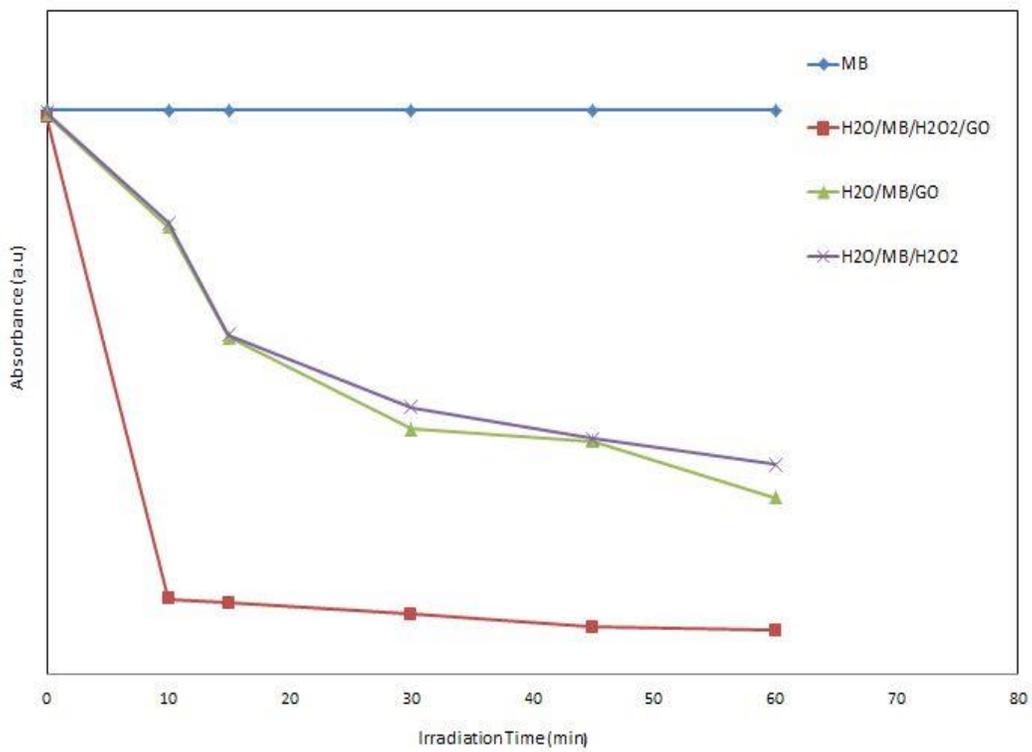

(b)

Fig 6: (a) ln($C_0/C_t$) versus Irradiation time and (b) Absorbance versus Irradiation time curves illustrating MB degradation by $H_2O/MB/GO$, $H_2O/MB/H_2O_2$, and $H_2O/MB/H_2O_2/GO$.

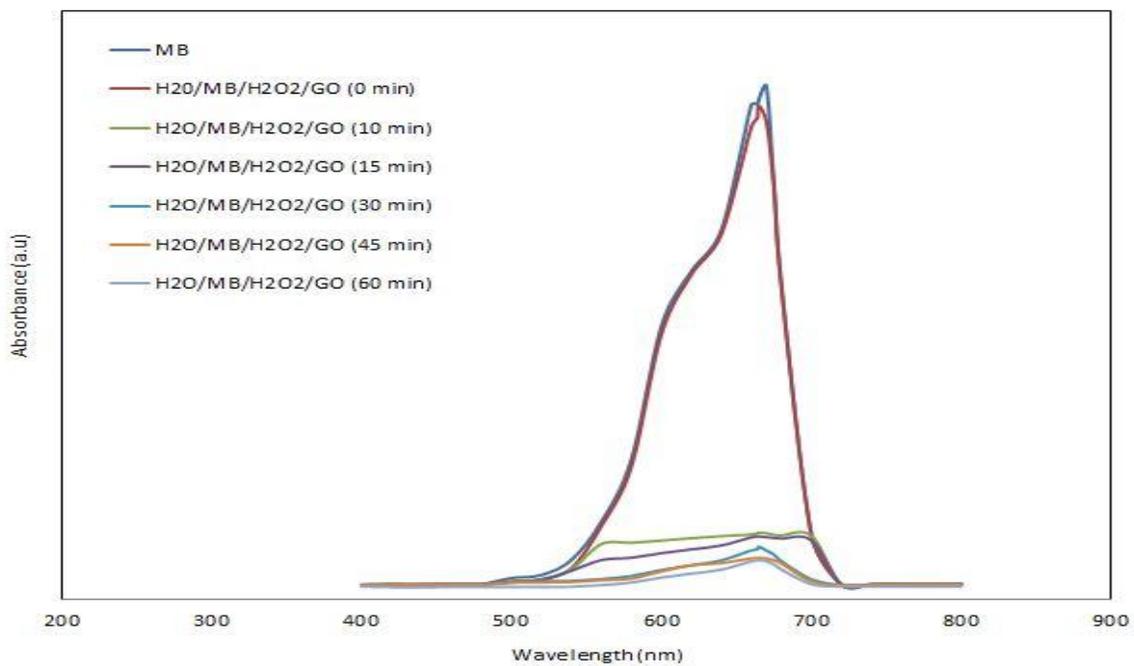

(a)

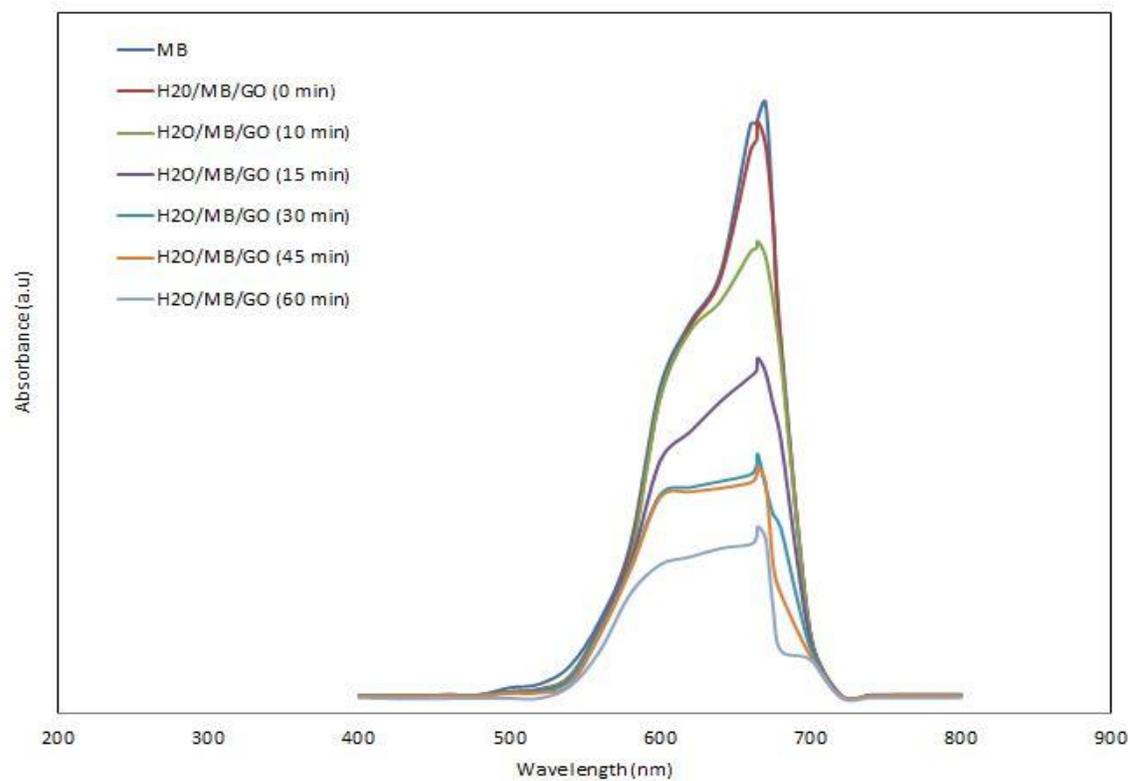

(b)

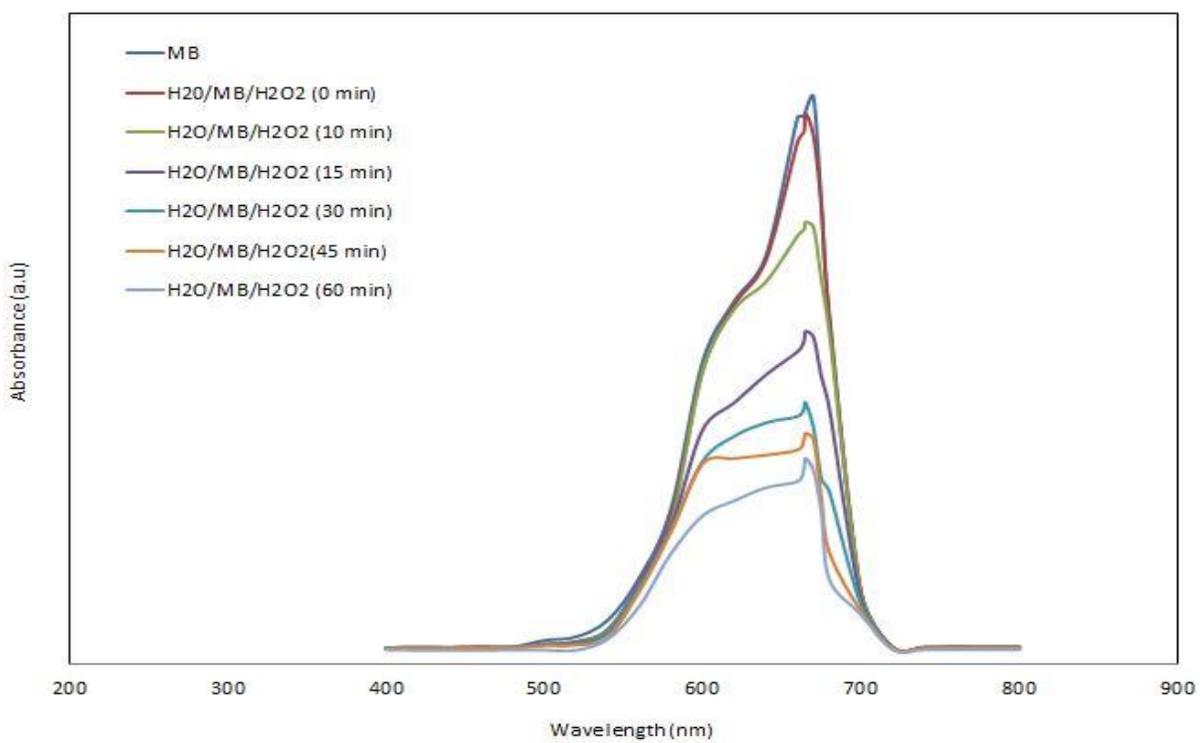

(c)

Fig 6: Time-dependent absorption spectra of MB solution during natural sunlight irradiation in the presence of (a) $H_2O/MB/H_2O_2/GO$, (b) $H_2O/MB/GO$, and (c) $H_2O/MB/H_2O_2$.

It is clear from Fig 6(a) and Fig 6(b) that, GO along with $H_2O_2$ as expected showed highest photocatalytic activity compared to that of $H_2O_2$. However, only $H_2O_2$ and only GO showed lower photocatalytic activity. GO nanoparticles is a catalyst for MB degradation and also $H_2O_2$ itself is a catalyst for MB degradation which takes 60 min for almost total degradation when both were used together.

The MB photo degradation was fitted to pseudo-first order kinetics by referring to the Langmuir-Hinshelwood kinetic model ((Equation (2)) [36,47]:

$$\ln(C_0/C_t) = kt \qquad (2)$$

Where $C_t$ is the concentration of MB at time, t, $C_0$ is the initial concentration of MB, and k is the pseudo-first order rate constant. The k value of respective concentrations was determined from knowing the values $C_t$ and was listed in Table 3. The correlation co-efficient ($R^2$) values are close to 1, which obeys the pseudo-first order kinetic model.

**Table 3. Degradation efficiency and pseudo-first order rate constant for photocatalytic degradation of MB by GO, $H_2O_2$, GO along with $H_2O_2$.**

| Samples | Concentration of MB (mM) | Degradation efficiency (%) | $R^2$ | Degradation Rate constant (min$^{-1}$) |
|---|---|---|---|---|
| $H_2O/MB/GO$ | 0.05 | 68.68% | 0.9297 | 0.01935 |
| $H_2O/MB/H_2O_2$ | 0.05 | 62.81% | 0.9064 | 0.01649 |
| $H_2O/MB/H_2O_2/GO$ | 0.05 | 92.23% | 0.8943 | 0.04258 |

As we can see, the absorbance versus irradiation time curves and $\ln(C_0/C_t)$ versus natural sunlight irradiation time curves

for MB photodegradation are non-linear because of the following probable reasons-
- Absorptivity co-efficient deviations occur when concentration is greater than 0.01mM and due to the electrostatic interactions between molecules in close proximity.
- Scattering of lights due to particulates in the sample.
- Chemical equilibrium shifting as a function of concentration.

The direct photolysis and the oxidative potential of $H_2O_2$ were proven to have a contribution on the degradation of MB. Notably, sunlight, GO and $H_2O_2$ together showed a marked effect. Increasing the DO concentration was beneficial for the photocatalytic degradation of MB. Correspondingly, the degradation rate constant increased with the DO concentration. For the sunlight/$H_2O$/MB/GO/$H_2O_2$ photocatalysis, $H_2O_2$ of lower dosage acted as electron acceptor to enhance the degradation efficiency. When the dosage was high, however, the degradation was suppressed due to the capture of •OH radicals and the competitive adsorption of $H_2O_2$. In order to abate the disadvantages caused by using a higher $H_2O_2$ dosage, sequential replenishment of $H_2O_2$ into sunlight/$H_2O$/MB/GO system was performed. Experimental results demonstrated that degradation efficiency was enhanced by the restraint of the capture of •OH radicals, the additional •OH radicals caused from the addition of $H_2O_2$, and the participation of oxygen in photocatalytic degradation [55].

It is evident that both degradation efficiency and degradation rate constant ($k_1$) increases remarkably with the increase of DO level for every system ((see Table 4, Table 5, Table 6 and Fig 7(a), 7(b) and 7(c)).

**Table 4. Degradation efficiency and rate constant for photocatalytic degradation of MB by $H_2O$/MB/$H_2O_2$/GO System with the variation of Dissolved Oxygen Concentration (DO).**

| Samples | Concentration of MB (mM) | DO (mgL$^{-1}$) | Degradation efficiency (%) | Degradation Rate constant (min$^{-1}$) |
|---|---|---|---|---|
| $H_2O$/MB/$H_2O_2$/GO | 0.05 | 2.8 | 87.4% | 0.035 |
| $H_2O$/MB/$H_2O_2$/GO | 0.05 | 3.1 | 91.3% | 0.041 |
| $H_2O$/MB/$H_2O_2$/GO | 0.05 | 3.5 | 92.2% | 0.043 |
| $H_2O$/MB/$H_2O_2$/GO | 0.05 | 3.9 | 97.6% | 0.062 |

**Table 5. Degradation efficiency and rate constant for photocatalytic degradation of MB by $H_2O$/MB/GO System with the variation of Dissolved Oxygen Concentration (DO).**

| Samples | Concentration of MB (mM) | DO (mgL$^{-1}$) | Degradation efficiency (%) | Degradation Rate constant (min$^{-1}$) |
|---|---|---|---|---|
| $H_2O$/MB/GO | 0.05 | 2.8 | 36% | 0.0074 |
| H$_2$O/MB/GO | 0.05 | 3.1 | 54% | 0.013 |
| H$_2$O/MB/GO | 0.05 | 3.5 | 68.7% | 0.019 |
| H$_2$O/MB/GO | 0.05 | 3.9 | 78% | 0.025 |

**Table 6. Degradation efficiency and rate constant for photocatalytic degradation of MB by H$_2$O/MB/H$_2$O$_2$ System with the variation of Dissolved Oxygen Concentration (DO).**

| Samples | Concentration of MB (mM) | DO (mgL$^{-1}$) | Degradation efficiency (%) | Degradation Rate constant (min$^{-1}$) |
|---|---|---|---|---|
| H$_2$O/MB/H$_2$O$_2$ | 0.05 | 2.8 | 18% | 0.003 |
| H$_2$O/MB/H$_2$O$_2$ | 0.05 | 3.1 | 46% | 0.01 |
| H$_2$O/MB/H$_2$O$_2$ | 0.05 | 3.5 | 62.8% | 0.017 |
| H$_2$O/MB/H$_2$O$_2$ | 0.05 | 3.9 | 75% | 0.023 |

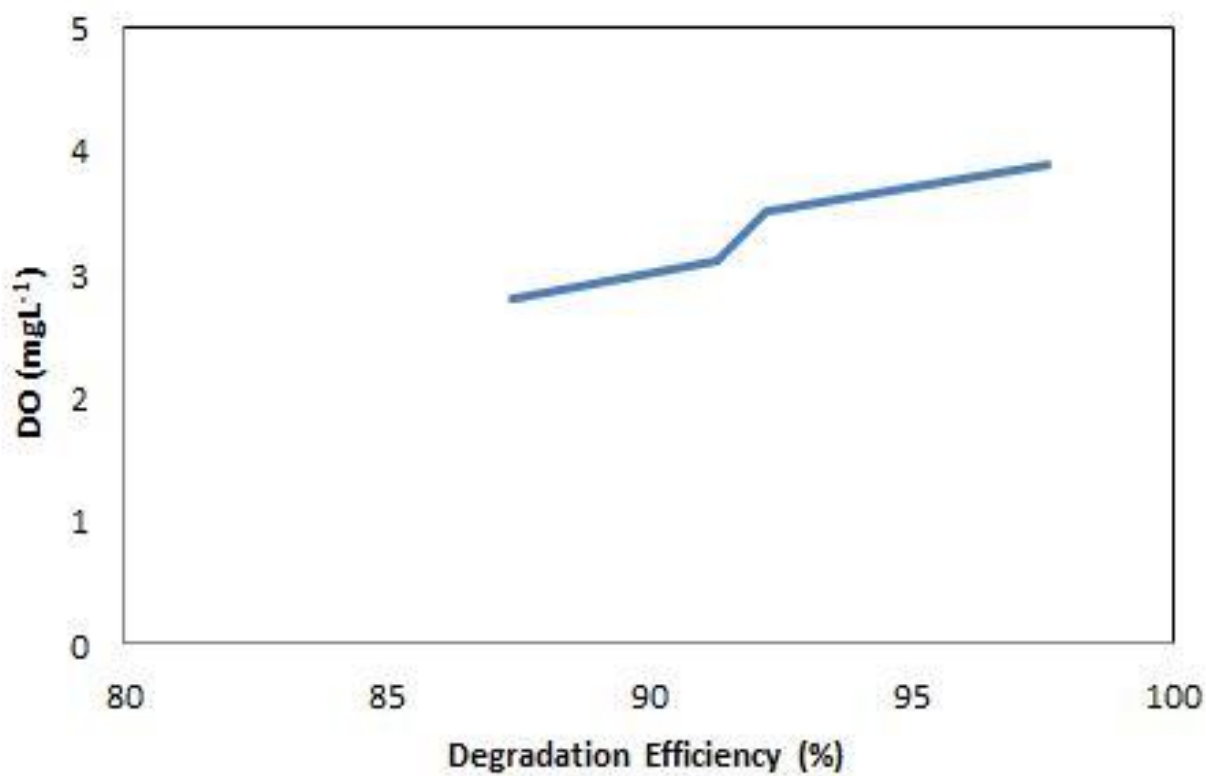

(a)

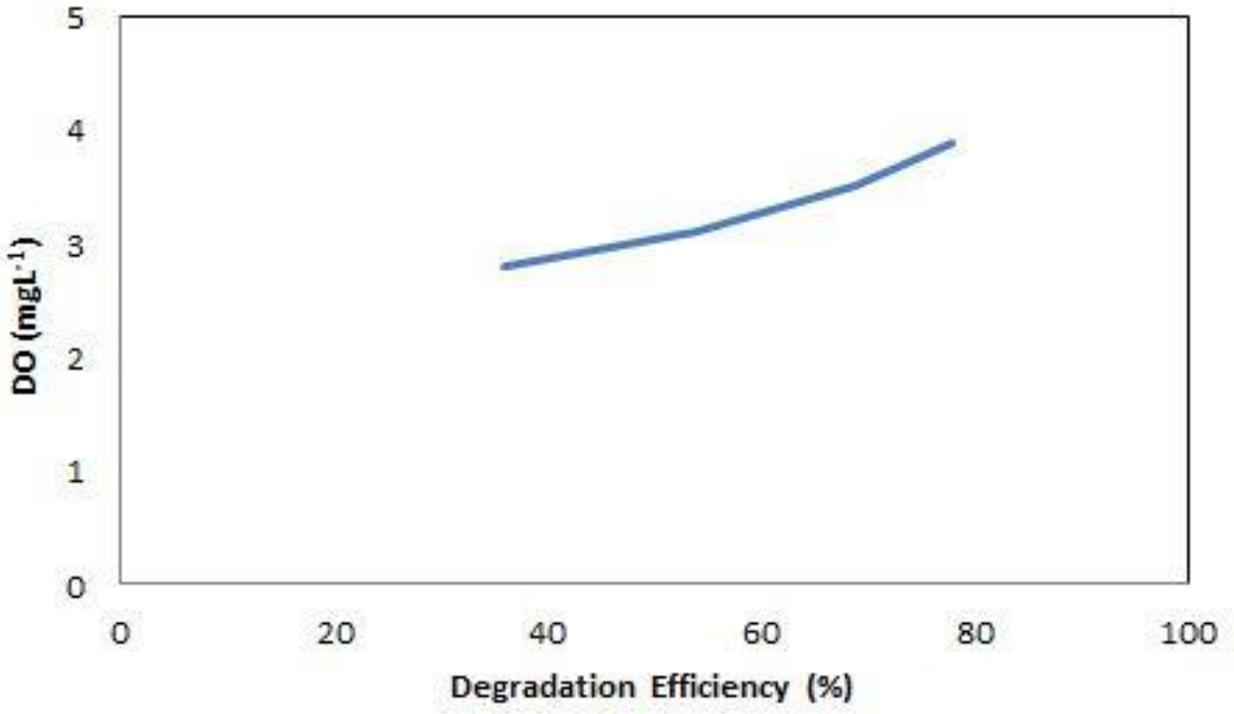

(b)

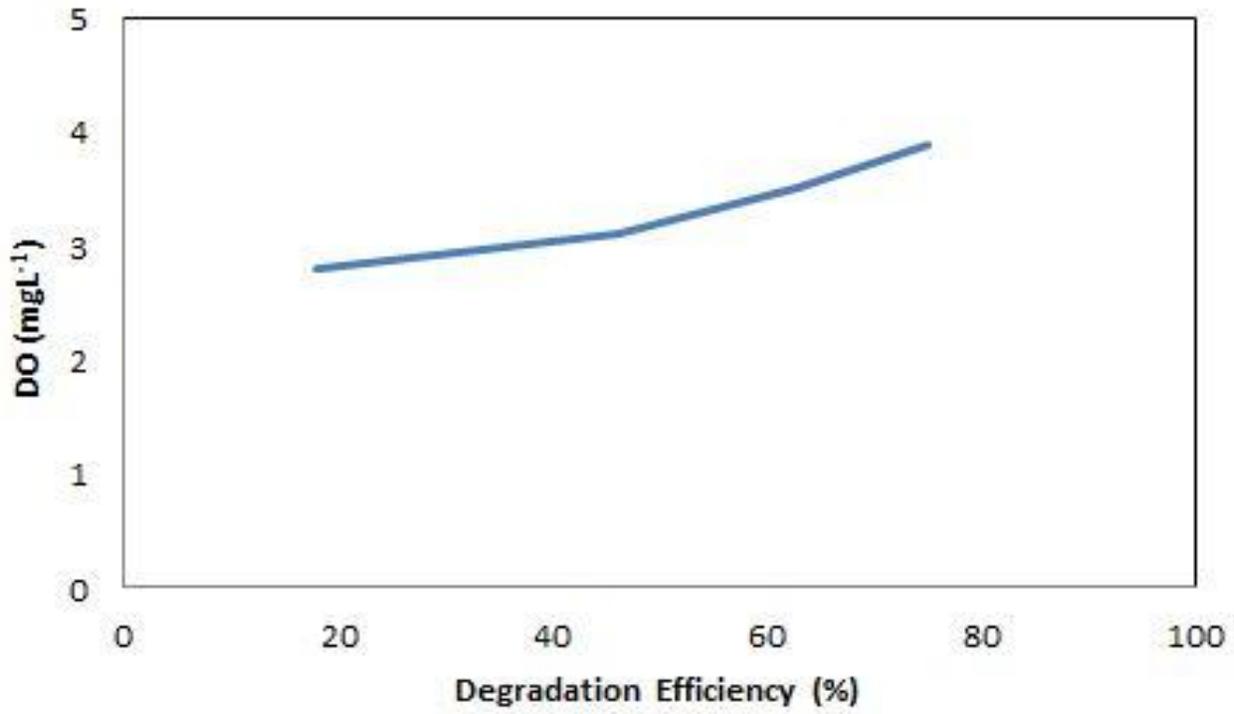

(c)

Fig 7: DO (mgL$^{-1}$) versus Degradation Efficiency (%) for (a) $H_2O$/MB/$H_2O_2$/GO, (b) $H_2O$/MB/GO and (c) $H_2O$/MB/$H_2O_2$ system.

*G. Mechanism*

The dye/$H_2O_2$/sunlight system involves the photocatalysis of hydrogen peroxide. The most accepted mechanism for this $H_2O_2$ photocatalysis is the rupture of the O-O bond by the action of sunlight forming two hydroxyl radicals (•OH) and these radicals in turns degraded MB.

$$H_2O_2 + e^- \rightarrow \bullet OH + OH^- \quad (3)$$
$$H_2O_2 + \bullet O_2^- \rightarrow \bullet OH + OH^- + O_2 \quad (4)$$
$$H_2O_2 + \bullet OH \rightarrow H_2O + \bullet OH_2 \quad (5)$$
$$HO_2 \bullet + \bullet OH \rightarrow H_2O + O_2 \quad (6)$$
$$H_2O_2 \rightarrow H_2O + \bullet O_2 \quad (7)$$

$$H_2O_2 \xrightarrow{h\nu} 2\bullet OH \quad (8)$$
$$\bullet OH + MB \rightarrow \text{Degraded Products} \quad (9)$$

The influence of $H_2O_2$ dosage on the degradation of MB can be explained in terms of the number of generated •OH radicals and the capture of •OH radicals [51-54]. It is well known that $H_2O_2$ can trap photoinduced $e^-$ to stabilize the paired $e^-$-$h^+$.

Additional •OH radicals could be yielded via the reaction between $H_2O_2$ and $e^-$ or $\bullet O_2^-$ ((eqs. (3) and (4)). As a result,

the addition of $H_2O_2$ into the photocatalytic system was expected to promote the degradation of MB. Exceeding the optimum dosage, however, the excess $H_2O_2$ would trap the •OH radicals to form weaker oxidant $HO_2\bullet$ radicals. Accordingly, the capture of ·OH radicals was occurred through ((eqn. (5) and (6)). The decline in the •OH radical concentration, trigged by the higher $H_2O_2$ dosage, restrained the degradation of MB. Correspondingly, the addition of $H_2O_2$ seemed to act as an oxygen source [55]. The mechanism of the photodegradation of MB in presence of GO only under natural sunlight irradiation can be described as follows:

$$GO + h\nu \rightarrow e_{CB}^- + h_{VB}^+ \quad (10)$$
$$V_o^{\bullet\bullet} + e_{CB}^- \rightarrow V_o^\bullet \quad (11)$$
$$V_o^\bullet + O_2 \rightarrow V_o^{\bullet\bullet} + \bullet O_2^- \quad (12)$$
$$e^-(\text{or } e_{CB}^-) + O_2 \rightarrow \bullet O_2^- \quad (13)$$
$$h_{VB}^+ + OH^- \rightarrow \bullet OH \quad (14)$$
$$e_{CB}^- + h_{VB}^+ \rightarrow \text{Heat} \quad (15)$$

A large amount of oxygen vacancies are present on GO surface. GO serve as an electron and hole source (from eq. 10) for degradation of organic dye; when GO nano materials are irradiated by natural sunlight with energy higher than or equal to the band gap of GO, an electron ($e_{CB}^-$) in the valence band (VB) can be excited to the conduction band (CB) with simultaneous generation of a hole ($h_{VB}^+$) in the VB. Oxygen vacancy defects ((see $V_o^\bullet$ and $V_o^{\bullet\bullet}$ in eqs. (11) and (12)) on the surface of GO act as a sink for the electrons and improve the separation of electron–hole pairs generated (in eq.9). The photoelectron can be easily trapped by electronic acceptors like adsorbed $O_2$, to further produce a superoxide radical anion ($\bullet O_2^-$) (in eq. 13). The photo induced holes can be easily trapped by OH⁻ to further produce a hydroxyl radical species (•OH) (in eq. 14). The generated superoxide anion radical (•$O_2^-$) and hydroxyl radical species (•OH) determine the overall photo catalytic reaction; for example, •OH is an extremely strong oxidant for the partial or complete mineralization of organic chemicals and/ or dyes like MB.

Since the band gap of GO was found as 3.26 eV [41] and when it is excited with an energy gap higher than the band gap energy, the electron and hole pairs will be the generated at the surface of GO. The defect sites in GO can act as trapping center for the excited carriers and thereby hinder the recombination process. MB molecule, which acts as an electron acceptor, readily accepts the photoexcited electrons resulting in the degradation of MB molecules. This is well supported with our results of UV-Vis spectra as shown in Fig 6(a), Fig 6(b), and Fig 6(c).

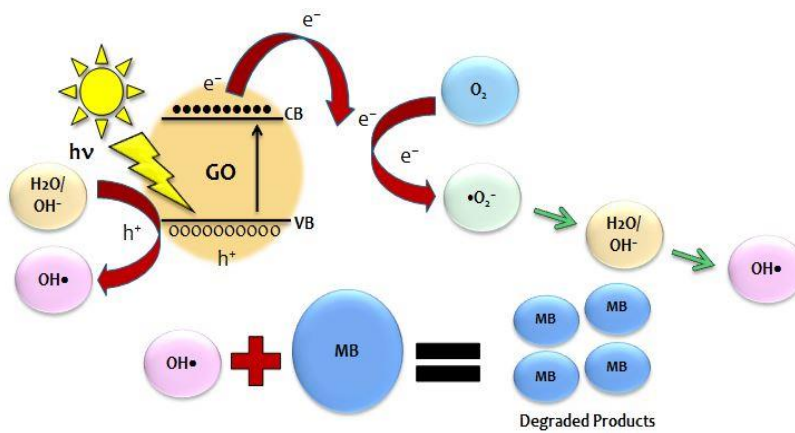

Fig 8: Mechanism of Direct Photocatalysis Using Graphene Oxide (GO)

In summary, GO nanostructures were synthesized by modified Hummer's method. XRD and FTIR studies reveal the existence of oxygenated functional groups in the GO. The degradation of MB by the GO nanostructures under sunlight irradiation was a pseudo first order reaction. The photo excited electrons from the surface state of GO under natural sunlight was responsible for the degradation of MB. Our experimental results demonstrated that GO nanostructures have promising applications in photocatalysis.

## CONCLUSION

Degradation of Methylene Blue under sunlight with GO nanoparticles as a catalyst takes around 60 min for almost total degradation when used with $H_2O_2$ as a positive catalyst. It can be used either alone or in combination with $H_2O_2$. $H_2O_2$ to activate the GO may also be used to speed up catalytic reactions for complete degradation. By increasing the quantity of GO, degradation time decreases under natural sunlight. GO and $H_2O_2$ can also be used individually for photo catalytic degradation of high concentration of Methylene Blue. Under natural sunlight irradiation GO along with $H_2O_2$ showed ~92% photodegradation efficiency after 60 min whereas only GO and only $H_2O_2$ showed ~69% and ~63% respectively. With the increase of initial concentration of dissolved oxygen (DO) from 2.8 to 3.9 mgL⁻¹, both degradation efficiency and rate constant increased markedly. Experimental study showed that the correlation co-efficient ($R^2$) values were close to 1, which obeyed the pseudo-first order kinetic model. The mechanism also described the whole photodegradation process in brief.

## Conflicts of Interests

There are no conflicts to declare.

## Acknowledgment

This work was supported by Department of Materials and Metallurgical Engineering (MME, BUET), The PP & PDC, BCSIR (Pilot Plant and Process Development Center, Bangladesh Council of Scientific and Industrial Research), Department of Glass and Ceramics Engineering (GCE, BUET) and Department of Chemistry, BUET.